\documentclass{article} 
\usepackage{iclr2026_conference,times}

\usepackage{amsmath,amsfonts,bm}

\def\eqref#1{equation~\ref{#1}}

\def\1{\bm{1}}

\DeclareMathAlphabet{\mathsfit}{\encodingdefault}{\sfdefault}{m}{sl}
\SetMathAlphabet{\mathsfit}{bold}{\encodingdefault}{\sfdefault}{bx}{n}

\usepackage{hyperref}
\usepackage{url}

\usepackage{iclr2026_conference,times}

\usepackage{latexsym}
\usepackage{amssymb}
\usepackage{amsmath}
\usepackage{bm}
\usepackage{microtype}
\usepackage{inconsolata}

\usepackage{graphicx}
\usepackage{booktabs}
\usepackage{multirow}
\usepackage{makecell}
\usepackage{siunitx}
\usepackage[table]{xcolor}
\usepackage{needspace} 
\usepackage{lineno}

\usepackage{enumitem}
\setlist[itemize]{leftmargin=1.5em, itemsep=1pt, topsep=1pt}
\usepackage{hyperref}
\usepackage{url}

\usepackage{caption}
\usepackage[normalem]{ulem} 

\definecolor{blue}{HTML}{006884}

\definecolor{myblue}{HTML}{006884}
\definecolor{or}{HTML}{D2691E}

\AtBeginDocument{
  \setlength{\abovedisplayskip}{10pt plus2pt minus5pt}
  \setlength{\belowdisplayshortskip}{8pt plus3pt minus3pt}
}

\title{Forecasting Future Language: Context Design for Mention Markets}

\author
{\textbf{Sumin Kim}$^{1,*}$,
    \textbf{Jihoon Kwon}$^{1,*}$,
    \textbf{Yoon Kim}$^{2}$,
    \textbf{Nicole Kagan}$^{3}$,\\
    \textbf{Raffi Khatchadourian}$^{4}$,
    \textbf{Wonbin Ahn}$^{5}$,
    \textbf{Alejandro Lopez-Lira}$^{6}$,\\
    \textbf{Jaewon Lee}$^{7}$,
    \textbf{Yoontae Hwang}$^{8}$,
    \textbf{Oscar Levy}$^{9}$,
    \textbf{Yongjae Lee}$^{10,**}$,
    \textbf{Chanyeol Choi}$^{1,**}$\\
    \\$^{1}$LinqAlpha \quad
    $^{2}$Massachusetts Institute of Technology \quad
    $^{3}$Kalshi \quad 
    $^{4}$IBM \quad
    $^{5}$LG AI Research \quad \\
    $^{6}$University of Florida\quad
    $^{7}$Seoul National University \quad
    $^{8}$Pusan National University \quad \\
    $^{9}$University of California, Berkeley\quad 
    $^{10}$UNIST \\
    [6pt]
    \footnotesize
    $^{*}$Equal Contribution. \quad $^{**}$Corresponding author.
}

\iclrfinalcopy

\begin{document}
\maketitle

\begin{abstract}
Mention markets, a type of prediction market in which contracts resolve based on whether a specified keyword is mentioned during a future public event, require accurate probabilistic forecasts of keyword-mention outcomes. While recent work shows that large language models (LLMs) can generate forecasts competitive with human forecasters, it remains unclear how input context should be designed to support accurate prediction. In this paper, we study this question through experiments on earnings-call mention markets, which require forecasting whether a company will mention a specified keyword during its upcoming call. We run controlled comparisons varying (i) which contextual information is provided (news and/or prior earnings-call transcripts) and (ii) how \textit{market probability}, (i.e., prediction market contract price) is used. We introduce Market-Conditioned Prompting (MCP), which explicitly treats the market-implied probability as a prior and instructs the LLM to update this prior using textual evidence, rather than re-predicting the base rate from scratch.
In our experiments, We find three insights: (1) richer context consistently improves forecasting performance;
(2) market-conditioned prompting (MCP), which treats the market probability as a prior and updates it using textual evidence, yields better-calibrated forecasts;
and (3) a mixture of the market probability and MCP (MixMCP) outperforms the market baseline. By dampening the LLM's posterior update with the market prior, MixMCP yields more robust predictions than either the market or the LLM alone.
\end{abstract}

\section{Introduction}
\label{sec:intro}

Prediction markets -- markets in which participants trade contracts on future events -- use contracts that pay a fixed payoff depending on the realized outcome.
As a result, contract prices are commonly interpreted as the probability of the outcome as collectively assessed by market participants~\citep{wolfers2004prediction, arrow2008promise}.
For example, a contract that pays \$1 if global oil prices rise by more than 10\% over the next quarter may trade at \$0.55, implying a 55\% market-implied probability.
Can probability assessments produced by prediction markets be improved by incorporating information from text, such as news and corporate disclosures, using modern NLP methods?

Recent work shows that large language models (LLMs) can serve as probabilistic forecasters for open-domain event questions, with probability estimates that are comparable to prediction-market forecasts~\citep{halawi2024approaching,alur2025aia}.
This naturally raises the question of whether LLMs and prediction markets can be complementary, with LLM-based signals helping to refine market-implied probabilities.
However, it remains unclear how LLM-based signals can be used to improve the predictive accuracy of market-implied probabilities, rather than competing with or replacing market forecasts.

In this paper, we first examine whether LLM-based signals can help improve market-implied forecasts in a specific class of prediction markets, rather than competing with them.
Specifically, we study mention markets—a class of prediction markets that resolve on whether a specified keyword appears verbatim in an upcoming earnings-call transcript.
Mention markets provide a particularly suitable testbed because forecasting their outcomes amounts to predicting future language, a task well aligned with LLMs’ language understanding capabilities, and because language used in earnings calls is closely linked to economically meaningful outcomes such as stock price reactions~\citep{price2012earnings,cicon2017say}.

Our proposed Market-Conditioned Prompting (\textsc{MCP}) method treats the market-implied probability(\textit{market probability}) as a prior and uses the model to update this prior based on external textual context.
By explicitly conditioning the model on the market probability, MCP elicits evidence-based revisions rather than independent forecasts.

Our contributions can be summarized as:
\begin{itemize} [topsep=0pt, parsep=0pt, partopsep=0pt]
    \item \textbf{Framework:} We formalize text-grounded forecasting as a \emph{market-conditioned updating} problem, where LLMs use textual context as evidence to revise the \textit{market probability}, rather than acting as standalone predictors.
    \item \textbf{Methodology:} We propose \textsc{MCP}, a prompting protocol that explicitly conditions LLMs on market prices, leading to significantly better calibration than standard prompting approaches.
    \item \textbf{Performance:} We show that a conservative mixture of the market prior and the MCP posterior (MixMCP) consistently outperforms the market baseline, confirming that LLMs can add value even to efficient markets.
\end{itemize}

\section{Background}
\begin{figure*}[t]
  \centering
  \includegraphics[width=1\textwidth]{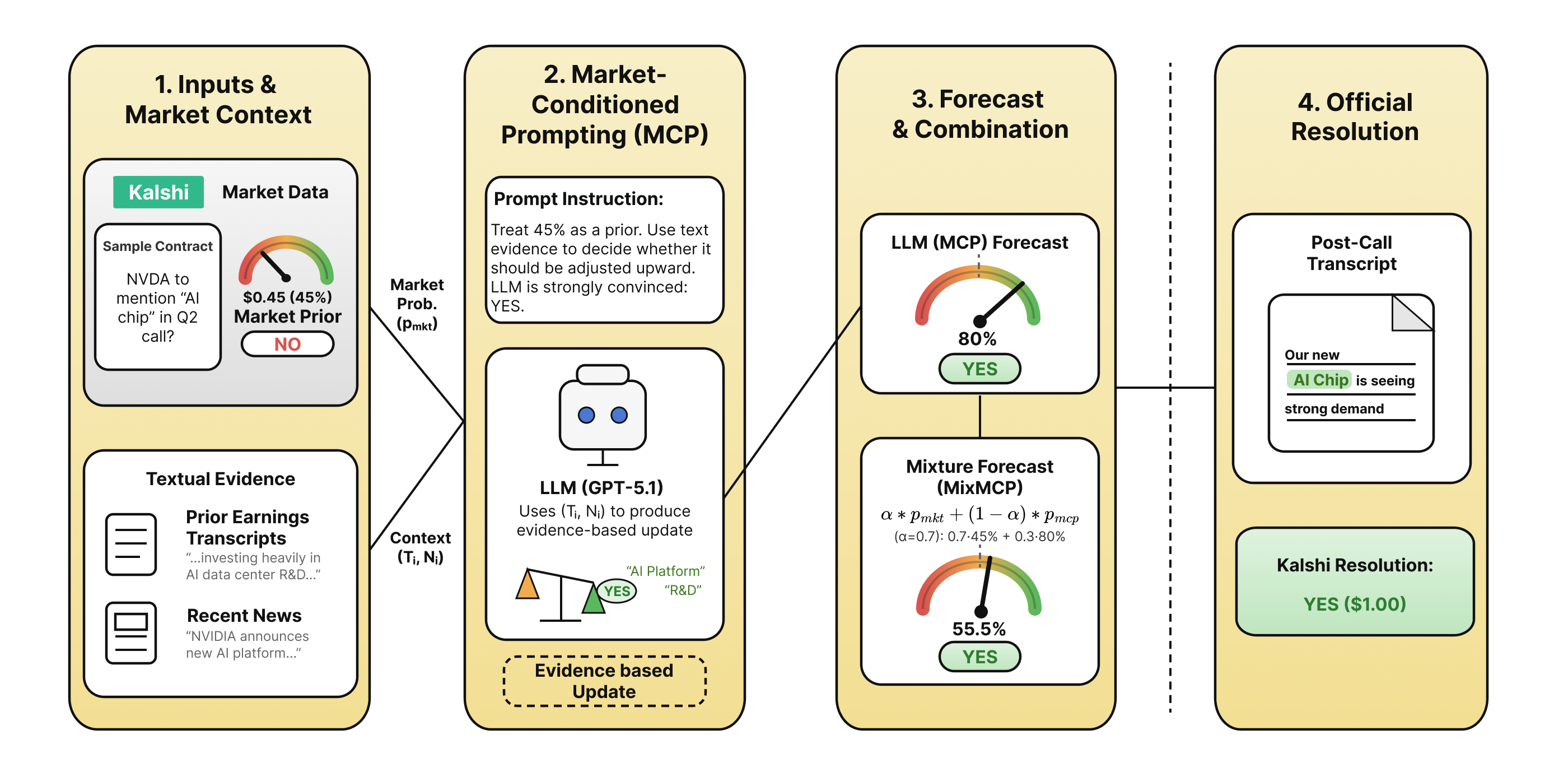}
\caption{
  \textbf{Resolving Market Ambiguity with \textsc{MixMCP}.}
  The diagram illustrates a mid-confidence scenario where the \textit{market probability} is uncertain (45\%).
  By conditioning on relevant textual evidence (e.g., earnings transcripts, news), \textsc{MCP} identifies a strong signal ($p_{\text{mcp}} = 80\%$) that the market has not fully priced in.
  Finally, \textsc{MixMCP} combines the stable market prior with the LLM's evidence-based update ($\alpha = 0.7$), producing a refined forecast (55.5\%) that correctly anticipates the resolution.
}
  \label{fig:motivational}

\end{figure*}
\paragraph{Earnings-Call Mention Markets.}
Kalshi operates prediction markets on earnings-call \emph{mentions}, where each contract resolves to \texttt{YES} if a specified keyword appears verbatim in the official post-call transcript.
A market instance $i$ is defined by an earnings event $e_i$, a target keyword $k_i$, and a pre-call cutoff time $t_i$.
At $t_i$, the YES-contract price reflects the \textit{market probability} $p_i^{\textsc{mkt}}$ that the keyword will be mentioned. As a text-grounded form of prediction market, earnings-call mention markets retain the standard interpretation of contract prices as implied probabilities, while shifting the prediction target to future language.
Because settlement is determined by exact string matching against the official post-call transcript, outcomes are objectively verifiable and evaluation is unambiguous.
Keywords are externally specified through Kalshi’s market-creation process, yielding precisely defined contracts and preventing researcher intervention in keyword selection.
This process also enforces standard safeguards against manipulation and ensures ethical, well-posed market design.

\paragraph{Forecasting Task.}
Given information available at $t_i$---including the market price and pre-call textual evidence such as news and prior earnings-call transcripts---a forecaster outputs a probability $p_i\in[0,1]$ for the binary outcome $y_i\in\{0,1\}$ determined by Kalshi's official resolution.

\paragraph{Evaluation Metrics.}
Probabilistic forecasts are commonly evaluated using the Brier score~\citep{brier1950,gneiting2007strictly},
\begin{equation}
\label{eq:brier}
\mathrm{Brier}=\frac{1}{N}\sum_{i=1}^{N}\left(p_i-y_i\right)^2,
\end{equation}
which measures the mean squared error between predicted probabilities and outcomes -- lower is better.
Calibration is assessed via Expected Calibration Error (ECE)~\citep{guo2017calibration}.
A well-calibrated forecaster should have actual accuracy match predicted confidence -- e.g., among predictions made with 70\% confidence, roughly 70\% should resolve to \texttt{YES}.
ECE quantifies this gap by partitioning predictions into $B$ equal-width bins $S_b=\{i: p_i\in(\frac{b-1}{B},\frac{b}{B}]\}$ and computing:
\begin{equation}
\label{eq:ece}
\mathrm{ECE}=\sum_{b=1}^{B}\frac{|S_b|}{N}\left|\bar{y}_b-\bar{p}_b\right|,
\end{equation}
where $\bar{y}_b$ and $\bar{p}_b$ are the mean outcome and mean predicted probability in bin $b$, respectively.
Lower ECE indicates better calibration. We use $B{=}10$ and also report accuracy and F1 score~\cite{chinchor1992muc4}.

\section{Methods} \label{sec:methods}
We study how to combine a strong market forecast with text based evidence available before the earnings call. The goal is to improve probabilistic accuracy without replacing the market prior.

\paragraph{Inputs and Base Forecaster.} For each market instance $i$, we take the \textit{market probability} at forecast time $p_i^{\mathrm{mkt}}$ and a prompt context $c_i$ built from available text. A fixed language model with parameters $\theta$ maps the context to a probability $p_i^{\mathrm{llm}} = f_{\theta}(c_i)$. We hold the model and decoding constant across all conditions.

\paragraph{Textual Context.} We build two source of pre-call evidence. The transcript block $T_i$ contains the prior quarter earnings call transcript for the same firm. The news block $N_i$ contains retrieved company related news articles~\citep{lewis2020rag} available before the cutoff $t_i$. We isolate the effect of each source by varying the context~\citep{liu2023prompting}
\begin{equation}
    c_i \in \{ \emptyset, N_i, T_i, (T_i, N_i) \}.
\end{equation}

\paragraph{Market Signal as Plain Text.}
A direct baseline inserts the market probability into the prompt without asking the model to treat it as a prior. Let $\phi(\cdot)$ be a deterministic textual rendering of the market probability. The baseline forecast is
\begin{equation}
    p_i^{\mathrm{llm}} = f_{\theta}(T_i, N_i, \phi(p_i^{\mathrm{mkt}})).
\end{equation}

This condition tests whether the model improves by simple exposure to the market number alone.

\paragraph{Market-Conditioned Prompting (MCP).}
We propose Market-Conditioned Prompting (MCP), a prompting protocol in which a large language model is explicitly provided with the \textit{market probability} together with pre-call textual evidence, and asked to output a revised probability forecast.

Formally, MCP produces a probability estimate of the form
\begin{equation}
\label{eq:mcp}
p_i^{\text{MCP}} = \text{LLM}_\theta(T_i, N_i \mid p_i^{\mathrm{mkt}}),
\end{equation} 
Here, $\text{LLM}_\theta(\cdot)$ denotes the fixed language model with parameters $\theta$, queried via our MCP prompt template (see Fig.~\ref{fig:mktcond}), which takes as input the market probability $p_i^{\mathrm{mkt}}$ (presented on a 0--100 scale in the prompt) together with the textual context $(T_i, N_i)$ and returns a single numeric probability parsed from the model’s output.

\paragraph{Mixture Forecast (\textsc{MixMCP}).} Even with an explicit prior, an LLM can overreact to weak or noisy cues. Following classic work on combining probability forecasts~\citep{genest1986combining,ranjan2010combining} and recent work on combining market baselines with LLM-based adjustments~\citep{alur2025aia}, we use a convex mixture that anchors on the market
\begin{equation}
    \label{eq:mix}
     p_i^{\text{mixMCP}} = \alpha \, p_i^{\text{mkt}} + (1 - \alpha) \, p_i^{\text{MCP}},
\end{equation}
Where $\alpha \in [0,1]$ is fixed across instances. This pooling rule keep the market forecast as a stable baseline while still allowing systematic improvements from evidence based updates.

\section{Experiments}
\label{sec:experiments}
In this section, we empirically evaluate the effectiveness of the methods defined in Section~\ref{sec:methods}.
In Sec.~\ref{subsec:exp_setup}, we describe the experimental setup, followed by Sec.~\ref{subsec:results} which demonstrates the predictive superiority of our proposed approach. 

\subsection{Experimental Setup}
\label{subsec:exp_setup}

\paragraph{Dataset.}
We evaluate on $N=856$ Kalshi earnings-call mention markets spanning 50 companies and 70 earnings events (Apr--Dec 2025).
Each instance $i$ has a cutoff time $t_i$ set to 7 days before the earnings call.
The label $y_i\in\{0,1\}$ is the Kalshi official resolution based on verbatim keyword presence in the released transcript (see Table~\ref{tab:appendix_dataset_example}).

\paragraph{Input Construction.}
At cutoff $t_i$, we collect two types of information.
For the market signal, we set $p_i^{\mathrm{mkt}}$ to the most recent Kalshi YES-contract price observed at or before $t_i$.
For textual evidence, $T_i$ is the prior-quarter earnings-call transcript, and $N_i$ contains up to 100 company-related news items retrieved via SERP API, each formatted with title, snippet, source, and date.

\paragraph{Implementation.}
All experiments use \texttt{GPT-5.1} with structured output on a 0--100 scale (rescaled to $[0,1]$), without fine-tuning.
We sweep $\alpha \in [0,1]$ and choose $\alpha = 0.7$ (with a plateau around $0.6$--$0.8$) on a held-out split.
Therefore, we set $\alpha=0.7$ for the mixture throughout.
To prevent leakage, forecasts are generated using only information available before $t_i$; the target earnings-call transcript is not observed at inference time.

\subsection{Results}
\label{subsec:results}

\paragraph{Our method, \textsc{MixMCP}, outperforms the \emph{market probability} baseline.}
As shown in Table~\ref{tab:mkt_vs_mix}, \textsc{MixMCP} combines a strong market forecast with model-based updates derived from market-conditioned prompting (MCP), achieving a lower Brier score and higher accuracy than the market alone.
In the following, we analyze why this combination is effective, examining the role of textual context, the impact of market-conditioned prompting, and the regimes in which these updates are most informative.

\begin{table}[t]
\caption{
Forecasting performance of \textsc{MixMCP} compared to the \textit{market probability} baseline. Arrows indicate the direction of better performance ($\downarrow$ lower is better, $\uparrow$ higher is better).}
\centering
\small
\setlength{\tabcolsep}{5pt}
{
\begin{tabular}{@{}l c c c c@{}}
\toprule
\textbf{Prediction} & \textbf{Brier} ($\downarrow$) & \textbf{ECE} ($\downarrow$) & \textbf{Acc} ($\uparrow$) & \textbf{F1} ($\uparrow$) \\
\midrule
\textsc{\textit{Market Probability}} & 0.1402 & \textbf{0.0651} & 79.8 & 0.840 \\
\textsc{mixMCP} & \textbf{0.1392} & 0.0666 & \textbf{80.3} &\textbf{ 0.842} \\
\bottomrule
\end{tabular}%
}
\label{tab:mkt_vs_mix}
\end{table}

\begin{table}[t]

\caption{
Prediction performance measured by Brier score, ECE, accuracy, and F1 score.
We compare the effect of providing different types of pre-event information as in-prompt context.
Note that \textsc{W/o Prompting} corresponds to the `Market Signal as Plain Text' baseline defined in Section~\ref{sec:methods}.}

\centering
\small
\setlength{\tabcolsep}{5pt}
{
\begin{tabular}{@{}l c c c c@{}}
\toprule
\textbf{Prediction} & \textbf{Brier} ($\downarrow$) & \textbf{ECE} ($\downarrow$) & \textbf{Acc} ($\uparrow$) & \textbf{F1} ($\uparrow$) \\

\midrule
\multicolumn{5}{l}{Textual context only} \\
\quad \textsc{Context} ($\emptyset$) & 0.2635 & 0.1993 & 63.1 & 0.707 \\
\quad \textsc{Context} ($N$) & 0.2372 & 0.1387 & 66.2 & 0.719 \\
\quad \textsc{Context} ($T$) & 0.2038 & 0.1105 & 70.0 & 0.733 \\
\quad \textsc{Context} ($T,N$) & 0.1991 & 0.0928 & 70.1 & 0.734 \\
\midrule
\multicolumn{5}{l}{Textual context with market signal} \\
\quad \textsc{\textbf{W/o Prompting}}$_{\{T,N,M\}}$ & \textbf{0.1674} & \textbf{0.0705} & \textbf{74.4} & \textbf{0.782} \\
\bottomrule
\end{tabular}
}
\label{tab:context}
\end{table}

\begin{table}[t]
\caption{
Effect of MCP.
We compare providing the \emph{market probability} as plain context with explicitly conditioning the model to reason about and update the market prior.
\textsc{W/o Prompting}$_{\{T,N,M\}}$ appends the \emph{market probability} as context, while \textsc{MCP} applies market-conditioned prompting.
}
\centering
\small
\setlength{\tabcolsep}{5pt}
{
\begin{tabular}{@{}l c c c c@{}}
\toprule
\textbf{Prediction} & \textbf{Brier} ($\downarrow$) & \textbf{ECE} ($\downarrow$) & \textbf{Acc} ($\uparrow$) & \textbf{F1} ($\uparrow$) \\
\midrule
\textsc{W/o Prompting}$_{\{T,N,M\}}$ & 0.1674 & 0.0705 & 74.4 & 0.782 \\
\textsc{\textbf{MCP}} & \textbf{0.1470} & \textbf{0.0514} & \textbf{78.2} & \textbf{0.822} \\
\bottomrule
\end{tabular}
}
\label{tab:prompting}
\end{table}

\paragraph{Richer context improves probabilistic forecasting performance.}
We first examine the effect of contextual information without any market-conditioned prompting, using a shared user input template across all variants (see ~\ref{fig:prompt_simple}).
As shown in Table~\ref{tab:context}, richer context consistently improves forecasting performance.
Prior-quarter transcripts yield larger gains than news alone, as they capture recurring themes and language patterns specific to each company's communication style, while combining transcripts and news provides complementary signals by reflecting both stable firm characteristics and recent developments. Providing the \emph{market probability} as an additional input---treated simply as another piece of context within the same template---further improves performance and yields the strongest results among all no-prompting variants.
This suggests that the market probability serves as a useful aggregate prior even when incorporated naively, without explicitly instructing the model to reason about or update it.

\paragraph{MCP improves over naive use of market probability.}
The results above indicate that the market probability can serve as a useful prior when provided as plain context.
We now examine whether explicitly instructing the model to reason about this prior leads to further improvements.
All variants share the same user input template (see.~\ref{fig:prompt_simple}) and differ only in how the market probability is incorporated.
Under \textsc{W/o Prompting}$_{\{T,N,M\}}$, the model receives transcripts, news, and the market probability as plain context, without any instruction to assess or revise the market signal.
As shown in Table~\ref{tab:prompting}, this naive inclusion underperforms the market baseline and results in poorer calibration. In contrast, Market-Conditioned Prompting (MCP) explicitly frames the market probability as a prior to be evaluated and updated (see.~\ref{fig:mktcond}).
With this instruction alone, MCP substantially improves calibration (ECE: 0.051 vs.\ 0.071), reduces Brier score (0.1470 vs.\ 0.1674), and improves accuracy (78.2 vs.\ 74.4) relative to \textsc{W/o Prompting}$_{\{T,N,M\}}$.
These results indicate that the gains arise not from access to the market probability itself, but from conditioning the model to reason about and revise it.

\begin{figure*}[t]
\centering

\captionsetup{font=small}
\captionof{table}{
Disagreement analysis between the \textit{market probability} baseline and \textsc{MCP}.
\textbf{Disagree $n$} denotes instances where the two methods differ in binary prediction.
Columns report the count of instances where each method achieves a lower Brier score.
}
\label{tab:disagreement}
\vspace{2pt}
\small
\setlength{\tabcolsep}{4pt}

\begin{tabular}{@{}l r rr@{}}
\toprule
\textbf{Market Prob. Bin} &
\textbf{Disagree $n$} &
\textsc{Market Prob.} &
\textsc{MCP} \\
\midrule
0--50\%        & 18 & 13 & 5  \\
\textbf{50--60\%} & 30 & 13 & \textbf{17} \\
\textbf{60--70\%} & 8  & 3  & \textbf{5}  \\
70--100\%      & 14 & 13 & 1  \\
\midrule
Total          & 70 & 42 & 28 \\
\bottomrule
\end{tabular}

\vspace{10pt}

\includegraphics[width=0.62\textwidth]{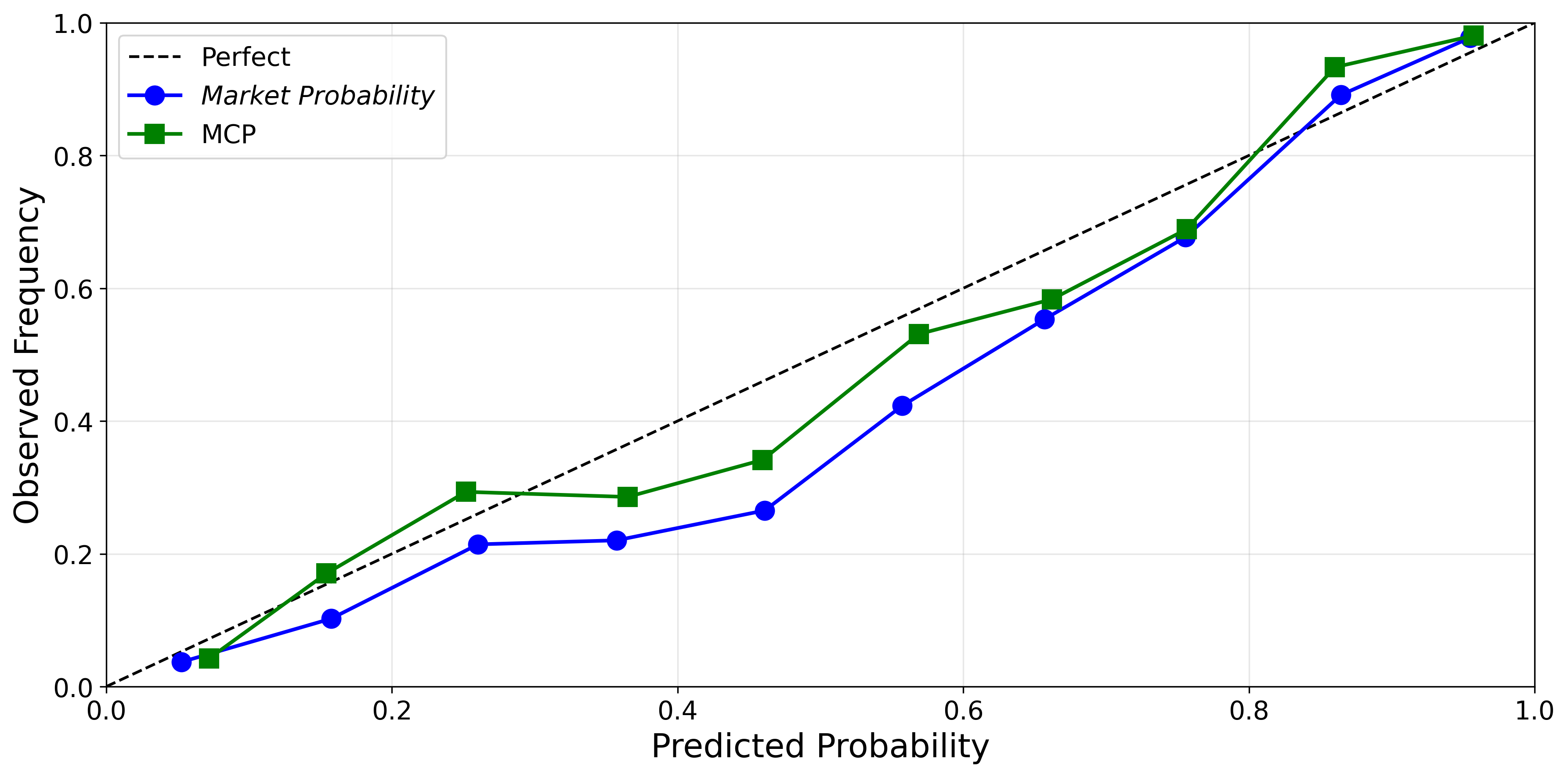}
\captionof{figure}{
Calibration plot comparing the \textit{market probability} baseline and \textsc{MCP}.
The x-axis shows predicted probability and the y-axis shows observed frequency of \texttt{YES} outcomes.
Points closer to the diagonal indicate better calibration.
}
\label{fig:reliability}

\vspace{10pt}

\captionof{table}{
Comparison of method variants.
\textsc{W/o Prompting} provides the \textit{market probability} as plain context,
while \textsc{MCP} instructs the model to reason relative to the market baseline.
\textsc{MixMCP} combines the \textit{market probability} and \textsc{MCP} with $\alpha=0.7$.
}
\label{tab:mcp_vs_mix}
\vspace{2pt}
\small
\setlength{\tabcolsep}{5pt}

\begin{tabular}{@{}l c c c c@{}}
\toprule
\textbf{Prediction} &
\textbf{Brier} ($\downarrow$) &
\textbf{ECE} ($\downarrow$) &
\textbf{Acc} ($\uparrow$) &
\textbf{F1} ($\uparrow$) \\
\midrule
\textsc{\textit{Market Probability}}
& 0.1402 & 0.0651 & 79.8 & 0.840 \\

\midrule
\textsc{MCP}
& 0.1470 & \textbf{0.0514} & 78.2 & 0.822 \\

\textsc{MixMCP}
& \textbf{0.1392} & 0.0666 & \textbf{80.3} & \textbf{0.842} \\

\bottomrule
\end{tabular}

\end{figure*}

\paragraph{MCP outperforms the market when market confidence is intermediate.}
To further understand the source of MCP’s gains, we analyze disagreement cases between MCP and the \emph{market probability}. Table~\ref{tab:disagreement} stratifies these disagreement cases by the market-implied probability and reports which method achieves a lower per-instance Brier score within each bin.
Among the 70 instances (8.2\% of the data) where the two methods differ in their binary predictions, MCP outperforms the market predominantly when the market signal is uncertain.
MCP demonstrates its strongest advantage in mid-confidence regimes, where the market signal is inherently ambiguous.
When market probabilities fall in the 50--60\% range, MCP achieves lower per-instance Brier scores in 17 out of 30 cases (56.7\%), and this advantage becomes even more pronounced in the 60--70\% range, where MCP wins 5 out of 8 cases (62.5\%).
These results show that MCP effectively leverages additional evidence to refine market forecasts precisely when the prior is least decisive.

Figure~\ref{fig:reliability} further supports this interpretation.
Across confidence bins, MCP improves overall calibration relative to the market (ECE: 0.051 vs.\ 0.065), with the largest gains occurring in the mid-confidence region.
At higher confidence levels, the market signal is already well-aligned with outcomes, and MCP largely preserves this behavior.
Taken together, these results indicate that MCP outperforms the market in mid-confidence cases, resolving ambiguity rather than overturning strong market beliefs.

\paragraph{\textsc{MixMCP}: Balanced Integration of Market and Textual Signals.}
While MCP provides informative corrections in mid-confidence regimes, it does not uniformly outperform the market across all scenarios. This suggests that while LLMs can extract novel insights from text, they may also introduce noise or over-interpret sparse cues. 
Motivated by this, we employ a convex mixture of MCP with the \emph{market probability}, MixMCP, to systematically balance the established market prior with the LLM’s posterior update.
As shown in Table~\ref{tab:mcp_vs_mix}, MixMCP achieves a lower Brier score (0.1392) compared to both the market (0.1441) and MCP alone (0.1470), while reaching the highest accuracy of 80.3\%.
In our experiments, an intermediate weighting ($\alpha = 0.7$) yielded the most robust results,
indicating that partially dampening the LLM signal improves predictive performance.

We interpret $\alpha$ as a parameter controlling the relative influence of the market signal and the LLM’s textual update. From an empirical perspective, $\alpha$ provides a simple mechanism for modulating how strongly the model relies on the market prior versus the LLM output. Viewed this way, the market acts as a stabilizing reference, while the LLM serves as a refinement component that adjusts predictions when additional semantic information is useful.

\section{Conclusion}
We study LLM forecasting on text-grounded mention markets and show that performance depends critically on context design.
Providing relevant textual evidence substantially improves performance over no-context baselines.
Treating the \textit{market probability} as a baseline to be evaluated  yields better-calibrated LLM forecasts, particularly by resolving ambiguity in mid-confidence regimes where market signals are uncertain.
Finally, combining \textit{market probability} and \textsc{MCP} via \textsc{MixMCP}  improves predictive performance, indicating that LLMs act as complementary refinements to market predictions.

\section*{Limitations}
We study a specific class of text-grounded markets (earnings-call keyword mentions), and results may not generalize to other contract types.
We evaluate offline using historical snapshots and do not study deployment dynamics or feedback effects.
LLM outputs remain prompt-sensitive; future work could explore robustness across models and additional calibration methods.

\bibliography{iclr2026_conference}
\bibliographystyle{iclr2026_conference}

\appendix
\onecolumn
\section{Appendix}

\subsection{Prompt Templates}
\label{app:prompts}

This appendix provides the prompt templates used in our experiments.
We include the user input template shared by all variants and the market-conditioned prompting instruction used in \textsc{MCP}.

\subsubsection{Shared User Input Template}

\begin{center}
\includegraphics[width=0.9\linewidth,height=0.85\textheight,keepaspectratio]
{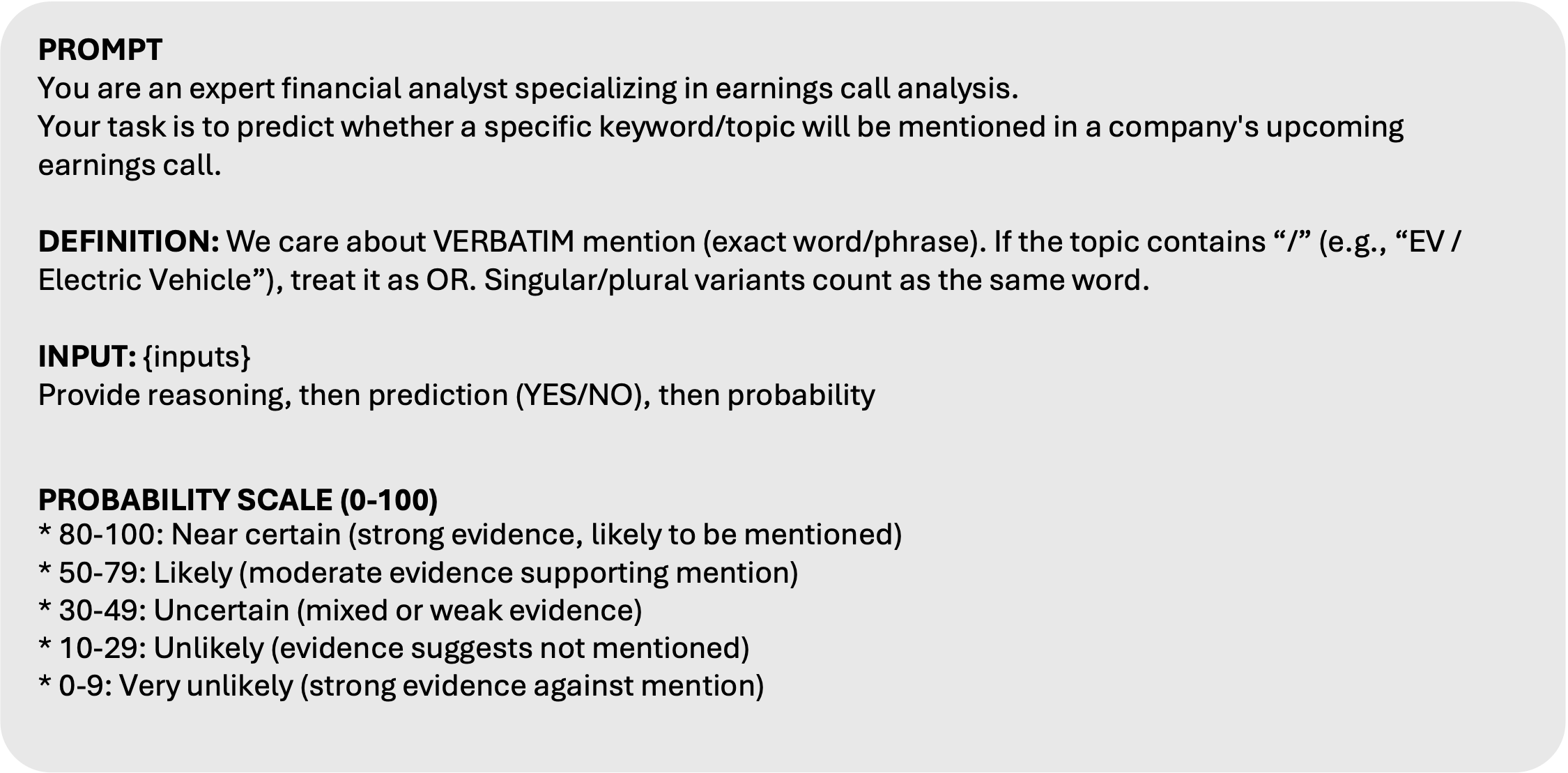}
\captionof{figure}{Shared user input template used across variants.}
\label{fig:prompt_simple}
\end{center}

\subsubsection{Market-Conditioned Prompting (\textsc{MCP})}

\begin{center}
\includegraphics[width=0.9\linewidth,height=0.85\textheight,keepaspectratio]
{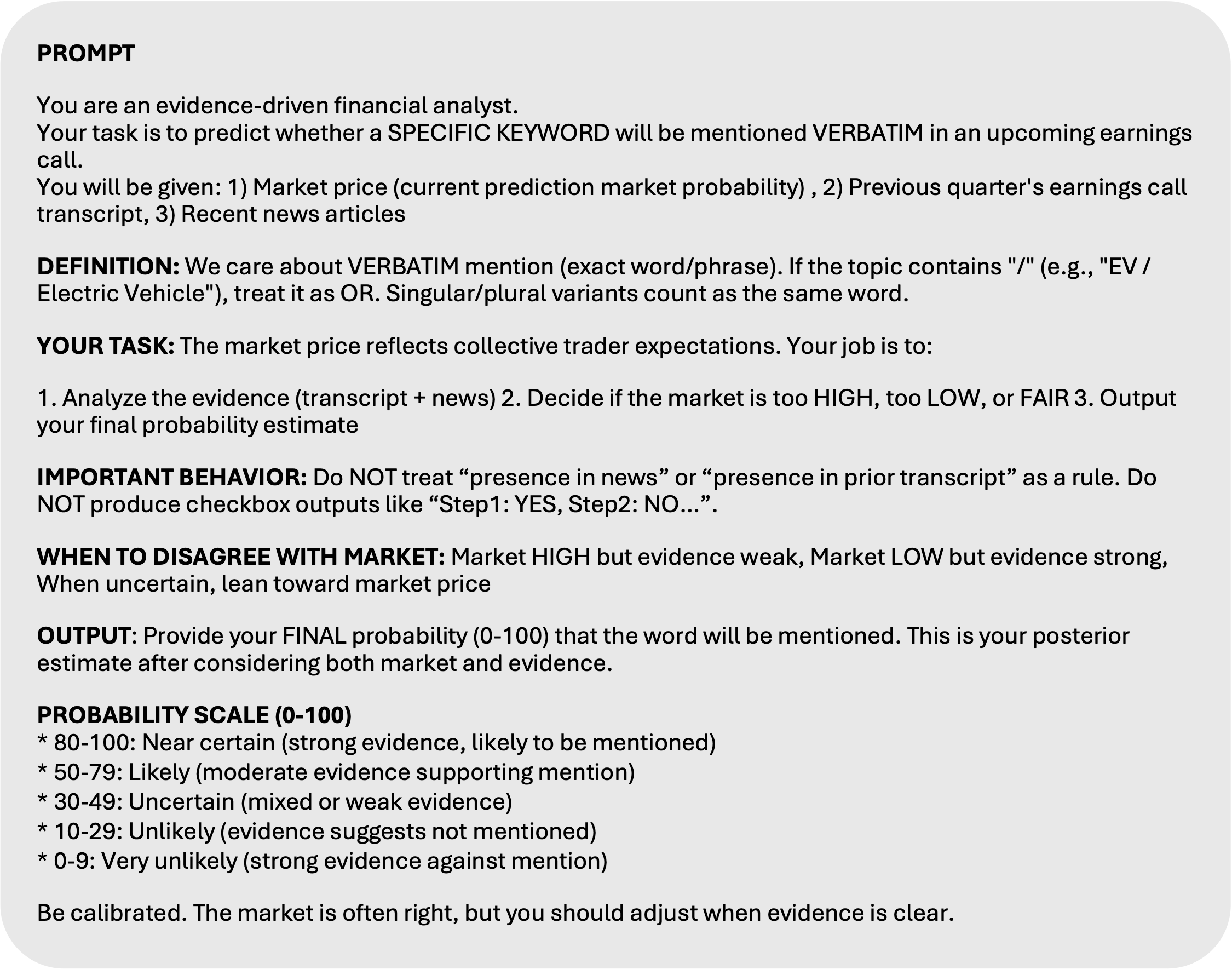}
\captionof{figure}{Market-conditioned prompting instruction used in \textsc{MCP}.}
\label{fig:mktcond}
\end{center}

\clearpage

\subsection{Dataset Examples}
\label{sec:appendix_dataset_examples}

\captionsetup{type=table}
\captionof{table}{
\textbf{Dataset example (illustrative).} Each row corresponds to a contract (event) with a short keyword summary.
\textit{Market Probability} reports the market-implied YES price (in \%) observed at a fixed horizon of $h=7$ days before the contract's resolution date.
\textit{Outcome} is the realized contract resolution (YES/NO).
}
\label{tab:appendix_dataset_example}

\begin{center}
\small
\setlength{\tabcolsep}{6pt}
\begin{tabular}{lllr l}
\toprule
Company & Resolution Date & Keyword & Market Prob. (\%) & Outcome \\
\midrule
AAPL & 2025-07-31 & iPhone     & 99 & YES \\
AAPL & 2025-07-31 & China      & 91 & YES \\
AAPL & 2025-07-31 & M3         & 47 & NO  \\
AAPL & 2025-07-31 & Streaming  & 36 & YES \\
NVDA & 2025-08-28 & AI         & 98 & YES \\
NVDA & 2025-08-28 & Blackwell  & 85 & YES \\
NVDA & 2025-08-28 & China      & 72 & YES \\
TSLA & 2025-07-22 & Robotaxi   & 67 & YES \\
TSLA & 2025-07-22 & Cybertruck & 82 & YES \\
TSLA & 2025-07-22 & FSD        & 91 & YES \\
NFLX & 2025-07-17 & Ad Tier    & 79 & YES \\
NFLX & 2025-07-17 & Squid Game & 73 & YES \\
AMZN & 2025-08-01 & AWS        & 99 & YES \\
AMZN & 2025-08-01 & Prime      & 95 & YES \\
AMZN & 2025-08-01 & Tariff     & 58 & YES \\
KO   & 2025-10-22 & China      & 65 & YES \\
KO   & 2025-10-22 & Tariff     & 42 & NO  \\
KO   & 2025-10-22 & Cane Sugar & 31 & NO  \\
\bottomrule
\end{tabular}
\end{center}

\end{document}